\documentclass[10pt,conference]{IEEEtran}
\makeatletter
\def\ps@headings{%
\def\@oddhead{\mbox{}\scriptsize\rightmark \hfil \thepage}%
\def\@evenhead{\scriptsize\thepage \hfil \leftmark\mbox{}}%
\def\@oddfoot{}%
\def\@evenfoot{}}
\makeatother
\usepackage{amssymb,color,amsmath}
\usepackage{paralist}
\usepackage{graphicx,epsfig}

\usepackage{graphicx}
\usepackage[urlcolor=rltblue,colorlinks=true]{hyperref} %
\definecolor{rltblue}{rgb}{0,0,0.75}
\usepackage{cite,manfnt}

\usepackage{algorithm2e}

\newcommand{\N}{\mathcal{N}}

\newtheorem{theorem}{\textbf{Theorem}}
\newtheorem{lemma}[theorem]{\textbf{Lemma}}
\newtheorem{definition}[theorem]{\textbf{Definition}}

\newcommand{\nix}[1]{}

\begin{document}
\title{\LARGE{ \huge\huge A Distributed Data Collection Algorithm for Wireless Sensor Networks with Persistent Storage Nodes}}
\author{
\authorblockN{Salah A. Aly$^{\dag\ddag}$~~~~~~~~ Ahmed Ali-Eldin$^{\S}$~~~~~~~~~H. Vincent Poor$^{\dag}$}
\authorblockA{$^{\dag}$Department of Electrical Engineering, Princeton University, Princeton, USA \\
$\ddag$Faculty of Computer Science, Modern Sciences and Arts University, MSA, October, Egypt\\
$^{\S}$School of Communication \& Information Technology,  Nile University, Smart Village, Egypt\\
%\{salah,poor\}@princeton.edu, ~~~~ahmed.alieldin@nileu.edu.eg
}

}
\maketitle

\begin{abstract}
A distributed data collection algorithm to accurately store and forward information obtained by  wireless sensor networks is proposed. The proposed algorithm does not depend on the sensor network topology, routing tables, or geographic locations of sensor nodes, but rather makes use of uniformly distributed   storage nodes. Analytical and   simulation results for this algorithm show that, with high probability,  the data disseminated by the sensor nodes can be precisely collected by querying any small set of storage nodes.
\end{abstract}
%%%%%%%%%%%%%%%%%%%%%%%%%%%%%%%%%%%%%%%%%%%%%%%%%%%%%%%%%%%%%%%%%%%%%%%%
\section{Introduction}\label{sec:intro}
Wireless sensor networks (WSNs) often consist of small devices (nodes)  with
limited processing ability,  bandwidth and power. They can be deployed in isolated or dangerous areas to monitor objects, temperatures, etc. or to detect fires, floods,  or  other incidents.  There has been extensive  research  on
sensor networks to improve their utility and efficiency~\cite{stojmenovic05}.

In this paper we consider a wireless sensor network $\N$ with $n$ nodes among which $k=n(1-\alpha)$ are sensing nodes  and $n-k$ are storage nodes, for small fractional $\alpha$ and $k/n\approx80\%$. The sensor and storage nodes
are distributed randomly in some region $\mathcal{R}$ and cannot maintain routing tables or shared knowledge of
network topology. Some nodes might disappear from the network due to
failure or battery depletion. It is of interest to design storage strategies
to collect sensed data from such sensors before they disappear
suddenly from the network.
Previous work on this problem has focused on situations in which either the network topology is known or the sensor nodes are able to maintain routing tables~\cite{goyeneche07,kamra06,lin07a}.

The authors in~\cite{aly09f,aly08e}  studied  distributed storage algorithms for wireless sensor networks in different topology in which $k$ sensor nodes (sources) want to disseminate their data to $n$ storage nodes with low computational complexity, where $k/n\approx20\%$. They used fountain codes and random walks on graphs to solve this problem. They also assumed that the total numbers of sources and storage nodes are not known. In  other words, they demonstrated an algorithm in which every node in a network can estimate the number of sources and storage nodes.
\nix{
\begin{figure}[t]
  \begin{center}
  \includegraphics[width=8.9cm,height=5.9cm]{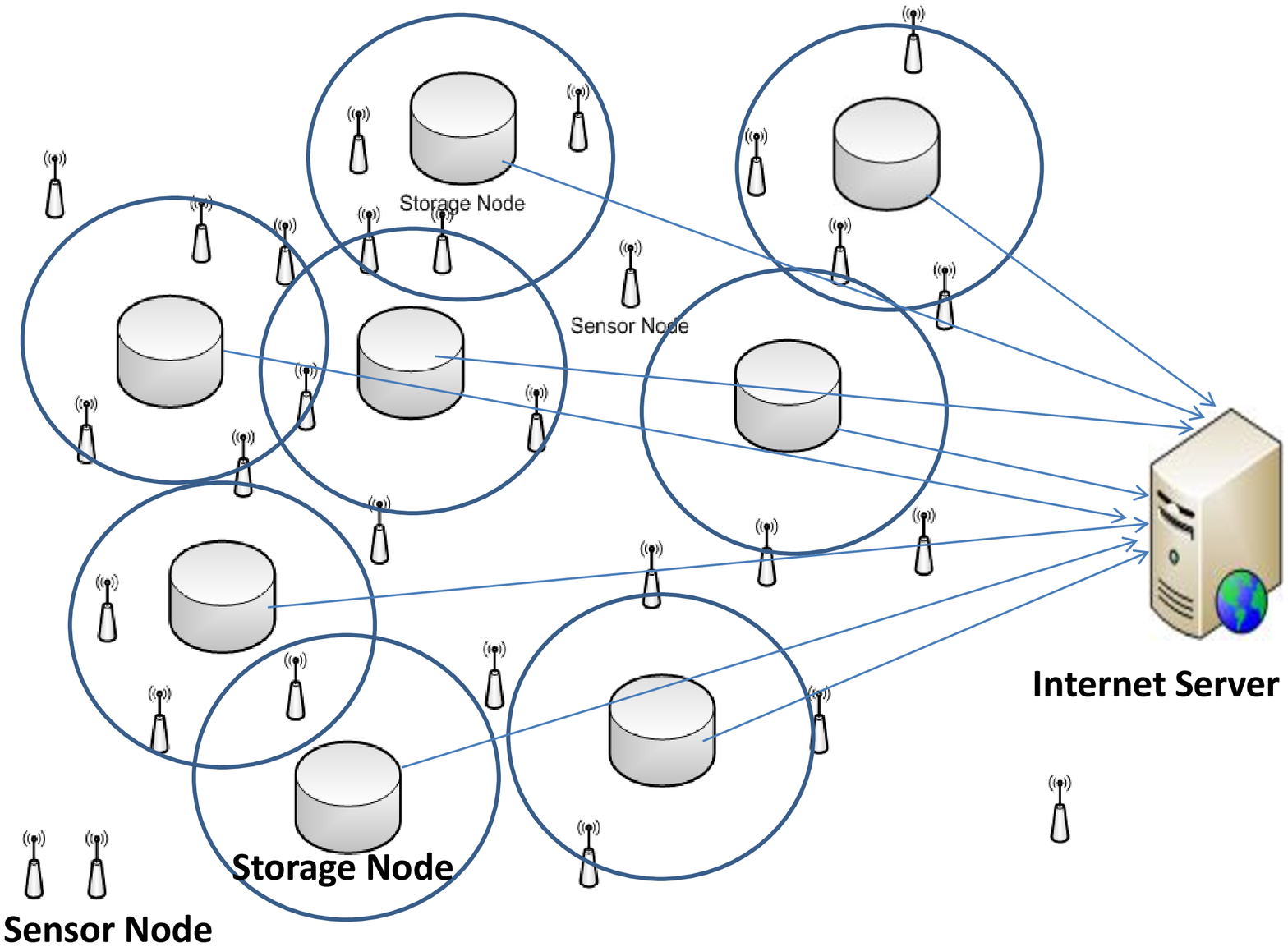}
  \caption{Network model representing  a wireless sensor network with sensing and storing nodes, and base stations nodes.}\label{fig:wsn-model}
  \end{center}
\end{figure}
}
In this work we solve the  storage problem in WSNs by developing data collection algorithms with persistent storage nodes and  dividing the region $\mathcal{R}$ into smaller regions.  We do not assume  routing or topology propositions about the network, as was done in~\cite{dimakis07,lin07a}. We consider situations in which the sensor nodes are distributed uniformly in $\mathcal{R}$, and, again, they do not maintain any routing tables or network topology.%, (see  Fig.~\ref{fig:wsn-model}).
There have been several clustering algorithms to aggregate nodes in wireless sensor networks. The most widely known are clustering by location or clustering using counters; see~\cite{liu07,stojmenovic05,younis2004distributed,bandyopadhyay2003energy,liu2007distributed} and references therein. The proposed data collection algorithm is suitable to use in terrains where we can not choose  positions of the sensor nodes or the cluster heads. In this case, the system is  self-stabilizing because if one node fails, no computations are needed to establish the cluster head.

The rest of the paper is organized as follows. In Section~\ref{sec:model} we present the network model and assumptions.  The distributed data collection algorithm is proposed in Section~\ref{sec:algDSA-I} and an analysis for this algorithm is presented in Section~\ref{sec:analysis}. In Section~\ref{sec:simulation}, we demonstrate performance and simulation results for the proposed algorithm. In  Section~\ref{sec:relatedwork}, we describe other work related to the proposed problem. Finally, the paper is concluded in Section~\ref{sec:conclusion}.

%%%%%%%%%%%%%%%%%%%%%%%%%%%%%%%%%%%%%%%%%%%%%%%%%%%%%%%%%%%%%%%%%%%%%%%%
\section{Network Model and Assumptions}\label{sec:model}
%\bigskip
Assume a large scale wireless sensor network with a set of sensing nodes and a set of storage nodes. Both are distributed randomly and uniformly in a given region $\mathcal{R}=L\times L$, where $L$ is the side length. The sensing nodes  have limited memory and bandwidth, and they might disappear from the network at any time due to limited battery lifetime. The storage nodes have large memory and bandwidth, but they do not sense information about the region.

We assume that the data collector (base station) is far away from the nodes \nix{as shown in Fig.~\ref{fig:wsn-model}}, but it is connected with a set of storage nodes. The sensor nodes are able to sense data and distribute it to the storage nodes.

\subsection{Assumptions}

We consider the following assumptions about the sensor network model $\N$:
\begin{compactenum}[i)]
\item Let $S=\{s_1,\ldots,s_k\}$ be a set of sensing nodes that are
    distributed randomly and uniformly in a given region $\mathcal{R}$. All sensor nodes have the same capabilities such as mobility, homogeneous, limited memory and power.
    \item Let $R=\{r_1,\ldots,r_{n-k}\}$ be the set
    of storage nodes such that $(n-k)/n=10\%\sim 20\%$. This assumption differentiates between the work     and  problem considered in~\cite{aly09f,aly08e,lin07a}. All storage nodes have the same amount of memory, power and bandwidth.
\item The nodes do not maintain routing or geographic tables, and  the
    topology of the wireless sensor network is not known.  Each  storage node $r_i$ can send
    multicasting  messages to  neighboring nodes. Also,
    each node $r_i$ can detect its total number of neighbors by sending a
    simple flooding query message, and any sensor node that responds  to this message
    will be a neighbor of this node. Therefore, our work is more general
    and different from the work done in~\cite{dimakis07,dimakis05} which depends on the knowledge of  network topology and routing tables. The
    degree $d_n(u)$ of a node $u$ is the total number of neighbors with a
    direct connection with this node.

\item  Each storage node has  a memory buffer of size $M$ and this buffer can be divided
    into smaller buffers, each of size $c$, such that $\epsilon=\lfloor M/c
    \rfloor$. For simplicity we assume that all storage nodes have equal memory size $M$.

\item Every node $s_i$ prepares a packet $packet_{s_i}$ with its $ID_{s_i}$,
    sensed data $x_{s_i}$,  and a \emph{flag} that is set to zero or one:
\begin{eqnarray}
packet_{s_i}(ID_{s_i},x_{s_i}, flag)
\end{eqnarray}
\item We will consider two different types of packets depending on the \emph{flag} value: initialization and
    update packets.  If the source node sends a packet and
    the \emph{flag} is set to zero, then it will be considered as an
    initialization packet. Otherwise, it will be considered as an update
    packet.
\item The network is divided into clusters (sub-regions). Every cluster region is identified by a storage node $r_i$, which exists in this cluster. Hence  the storage node is also called the cluster head. Every storage node will accept the incoming packets with probability one, and will update its buffer if the flag is set to one.
\end{compactenum}

\subsection{Distance Measurement and Clusters Distribution}
Since the sensing and storage nodes are distributed randomly, the distances between nodes are not known, but can be measured using the coverage radius of the nodes. When a storage node sends a flooding beacon message to all other sensing nodes,  those sensing nodes that can receive this beacon  will respond with reply  messages. The storage node will accept these reply messages and decide to receive  information from a node $s_i$ based on the   following comparison:
\begin{eqnarray}\label{eq:distance-sr}
d_{r_is_j}\leq \delta,
\end{eqnarray}
where  $d_{r_is_j}$ denotes the (Euclidean) distance between $r_i$ and $s_j$, and $\delta$ is a fixed distance for all storage nodes. In this case if the distance $d_{r_is_j}$ is greater than $\delta$, then the sensing node $s_j$ does not lie in the cluster identified by $r_i$~~\cite{liu07}. \nix{Fig.~\ref{fig:wsn-model} shows the coverage area of the sensor and storage nodes distributed randomly in the area $\mathcal{R}$~\cite{liu07}.}

%%%%%%%%%%%%%%%%%%%%%%%%%%%%%%%%%%%%%%%%%%%%%%%%%%%%%%%%%%%%%%%%%%%%%%%%
\section{Distributed Data Collection Algorithms}\label{sec:algDSA-I}
In this section, we propose a distributed data collection algorithm for the storage problem proposed in the previous section.
The clustering storage algorithm runs  in  the following phases:

\begin{algorithm}[t!]
\SetLine%
\KwIn{A sensor network with $S=\{s_1,\ldots,s_k\}$ source nodes,  $k$ source packets $x_{s_i},\ldots,x_{s_k}$, and $n-k$ storage nodes $R=\{r_1,r_2,...,r_{n-k}\}$.}%
\KwOut{storage buffers $y_1,y_2,\ldots,y_{n-k}$ for all storage nodes $R$.} %
\ForEach{storage node $r_i=1:n-k$}%
    {
    Generate a beacon packet with its $ID_{r_i}$ and send flooding message to all sensing neighbors\;%
    Every sensing node will decide the storage nodes to connect to\;
    }%

\ForEach{source node $s_i, i=1:k$}%
    {
    Generate header of $x_{s_i}$ and $flag=0$\;%
    Prepare the $packet_{s_i}$\;%
    Send the $packet_{s_i}$ to storage nodes\;%
    }%

\While{source packets remaining}%
    {
    \ForEach{node $r_j$ receives packets}%
        {

        \If{the flag=0}%
                {
                Put $x_{s_i}$ into $r_j$'s buffer\;%
                }
            \Else
                {Update the $y_j$ buffer of the storage node $r_j$
                 $y_j=y_j \oplus x_{s_i}$\;
                }%

               }%
        }%

\mbox{}\\%x

\caption{DSA-I Algorithm: Distributed data collection algorithm  for a WSN in which
the data is disseminated using multicasting messages to all storage nodes.} \label{alg:DSA-I}
\end{algorithm}

\begin{enumerate}
  \item[i)] \textbf{Clustering phase:}
  We assume that the sensor network has $k/n\approx 80\%$ sensing nodes, and $(n-k)/n \approx 20\%$ storage nodes. All clusters in the network are established using clustering algorithms~\cite{liu07,stojmenovic05}. In the clustering phase, each storage node sends a flooding beacon message with its ID to all neighboring nodes in the network. Due to the random locations of the sensing nodes, some nodes will be able to receive this message and reply with their IDs to the storage nodes. In addition the sensing nodes will store the IDs of the storage nodes in which they received beacon messages:
   \begin{eqnarray}
      packet_{r_i \rightarrow S}(ID_{r_i})
      \end{eqnarray}

  \item [ii)] \textbf{Sensing phase:} In the sensing phase, the sensor nodes sense data from the environment. Once the data is collected, they send their packets to the storage node, from which they have received beacon packets:
      \begin{eqnarray}
      packet_{s_i \rightarrow R_{s_i}}(ID_{s_i},x_{s_i},R_{s_i}, flag),
      \end{eqnarray}
      where $R_{s_i}$ is the set of storage nodes with whom $s_i$ is connected.     The flag value determines whether the packet contains an update or initially sensed data. The update data from the sensing nodes will occur whenever they sense new information about the surrounding environment.

  \item [iii)] \textbf{Data collection and storage phase:} When a sensing node senses the environment, it sends its packets to its storage nodes. The storage nodes collect the incoming packets and store them encoded in their own buffer. Based on the type of the incoming packets, the storage nodes will store these packets or update the existing data in their buffers.
        \item [iv)] \textbf{Querying phase:} The query process can be done by the base station or server that collects all data from the storage nodes. In the following sections we will study the total number of nodes that must be queried in order to obtain the data sensed  by the sensor nodes.

\end{enumerate}

%%%%%%%%%%%%%%%%%%%%%%%%%%%%%%%%%%%%%%%%%%%%%%%%%%%%%%%%%%%%%%%%%%%%%%%%
\section{DSA-I Analysis}\label{sec:analysis}
In this section we will analyze the proposed data collection algorithm, which we call DSA-I.
\begin{lemma}
With  high probability, the data collector can retrieve information about the sensing nodes if
\begin{equation}\epsilon \geq k/(n-k), \end{equation}
where $\epsilon$ is the number of buffers in each storage node.
\end{lemma}
%\medskip

\begin{lemma}\label{lem:1senser1storage} The probability that  a sensor $s_j$ lands in the range of a storage node $r_i$ is given by
\begin{equation}
\frac{\pi\delta ^2 -a}{L^2 k}.
\end{equation}
\end{lemma}
\begin{proof}
We know that the sensor and storage nodes are distributed independently and uniformly in the region $\mathcal{R}$.  So the probabilities that a randomly chosen sensor is $s_j$ and a randomly chosen storage node is $r_i$ are given by
\begin{eqnarray}
\Pr(s_j)=\frac{1}{k} \mbox{~~~ and~~~}
\Pr(r_i)=\frac{1}{n-k}.
\label{eq:prob1}
\end{eqnarray}

Let us define the random variable $X_{r_is_j}$ to  indicate the event that one of the sensors $s_j$ lies within a radio range $\delta$ of a storage node $r_i$, for $1\leq j \leq k$ and $1 \leq i \leq n-k$:
\begin{eqnarray}
X_{r_is_j}=\left\{
  \begin{array}{ll}
    1, & \hbox{if $d_{r_is_j} \leq \delta$;} \\
    0, & \hbox{if $d_{r_is_j} > \delta$.}
  \end{array}
\right.
\end{eqnarray}
where  $d_{r_is_j}$ is defined in (~\ref{eq:distance-sr}).
We also define the random variable $Y_{r_is_j}$ to indicate  the probability that any of the sensor nodes $s_j$ lies within the range of a given storage node $r_i$, so:
\begin{equation}
\Pr(Y_{r_is_j}=1)=\frac{\pi\delta ^2 -a}{L^2}
\label{eq:SenStor}
\end{equation}
 which is the area covered by the radio range within the region $\mathcal{R}$ divided by the total area of $\mathcal{R}$, and $a$ is the area of the portion of the radio range of the storage node that falls outside $\mathcal{R}$.
  The previous terms are obtained assuming a uniform probability distribution, therefore, the probability that a particular sensor $s_j$ lies within the radio range of a storage node $r_i$ is obtained by multiplying $\Pr(s_j)$ in~(\ref{eq:prob1}) by~(\ref{eq:SenStor}),  so, \begin{equation}
\Pr(X_{r_is_j}=1)=\frac{\pi\delta ^2 -a}{L^2 k}.
\end{equation}
\end{proof}
%%%%%%%%%%%%%%%%%%%%%%%%%%%%%%%

The following lemma follows from Lemma~\ref{lem:1senser1storage}, and its proof is a direct consequence.
\begin{lemma}
 The probability that a sensor $s_j$ lands in the range of all storage nodes $R$ is given by
\begin{eqnarray}
\Big(\frac{\pi\delta ^2 -a}{L^2 k}\Big)^{n-k}.
\end{eqnarray}
\end{lemma}

\begin{proof}
By Lemma~\ref{lem:1senser1storage}, we know that the probability of one sensor located in one storage node is given by
\begin{eqnarray*}
P(X_{d_{r_is_j}=1})=\frac{\pi\delta ^2 -a}{L^2 k}.
\end{eqnarray*}
Since all storage nodes are distributed randomly and uniformly in the region, then we have
\begin{eqnarray*}
f(X_{d_{R~s_j}=1})&=&\prod_{i=1}^{n-k} P(X_{d_{r_is_j}\leq \delta})\\
&=&\Big(\frac{\pi\delta ^2 -a}{L^2 k}\Big)^{n-k}.
\end{eqnarray*}
\end{proof}
We also turn our attention to study the probability of all sensor nodes sited at one particular storage node $r_i$. In this case, the coefficients $p_{i1},p_{i2},p_{i3},\ldots,p_{ik}$ of the $i^{th}$ in the storage code $C$ are not zeros. In the other words,   $p_{ij} \neq 0$ for all $j=1,2,\ldots,k$.

\begin{lemma}
 The probability of all sensors $S$ sited in the range of a storage nodes $r_i$ is given by
\begin{eqnarray}
\Big(\frac{\pi\delta ^2 -a}{L^2 k}\Big)^{k}.
\end{eqnarray}
\end{lemma}
%%%%%%%%%%%%%%%%%%%%%%%%%%%%%%%%%%%%%%%%%%%%%%%%%%%%%%%%%%%%%%%%%%%%%%%%
\section{Performance and Simulation Results}\label{sec:simulation}
In this section, we study  the performance of the proposed algorithm for
WSNs through simulation. The main performance
metric we investigate is the successful decoding probability versus the query ratio. We assume a square region $\mathcal{R}$ of size $L \times L$ in the plane, in which $L=100$.   Recall that, a  sensor node lies in the coverage radius of a storage node if
$d_{r_i,s_j} \leq \delta$,
in which $\delta$ is covering radius of the storage nodes.
%\vspace{+0.1in}%
\begin{definition}(Storage Nodes Query Ratio)
 Let  $h$ be the number of storage nodes that are queried among the $n'=n-k$ storage nodes  in $\mathcal{R}$. Let $\eta$ be the ratio between the number of queried nodes and the
number of storage nodes $n'$, i.e.,
\begin{eqnarray}\label{eq:eta}
\eta=\frac{h}{n'}.
\end{eqnarray}
\end{definition}
%\vspace{+0.1in}%
\begin{definition}(Revealed Sensors Ratio)
 We  define the ratio of the number of sensor nodes $k^\prime$, in which their data is retrieved based on querying $h$ storage nodes, to the total number of sensor nodes $k$ as the \emph{revealed sensors ratio} $\rho$:
\begin{equation}
\rho=k'/k.
\end{equation}
\end{definition}
\begin{definition}(Successful Decoding Probability)
The \emph{successful decoding probability} $P_s$ is the probability that the $k$
source packets are all recovered from the $h$ querying storage nodes.
\end{definition}
\vspace{+0.1in}%

The main  metric that we investigate is the revealed sensors ratio.  It shows the amount of information that we successfully are  able to obtain based on the proposed algorithm. We study the relationship between the range of the storage nodes $\eta$ and the revealed sensors ratio $\rho$. We first  fix $\eta$ and change the ratio between the range of the storage nodes and the region length $L$.

\begin{figure}[t]
  \begin{center}
 \includegraphics[scale=0.33,angle=270]{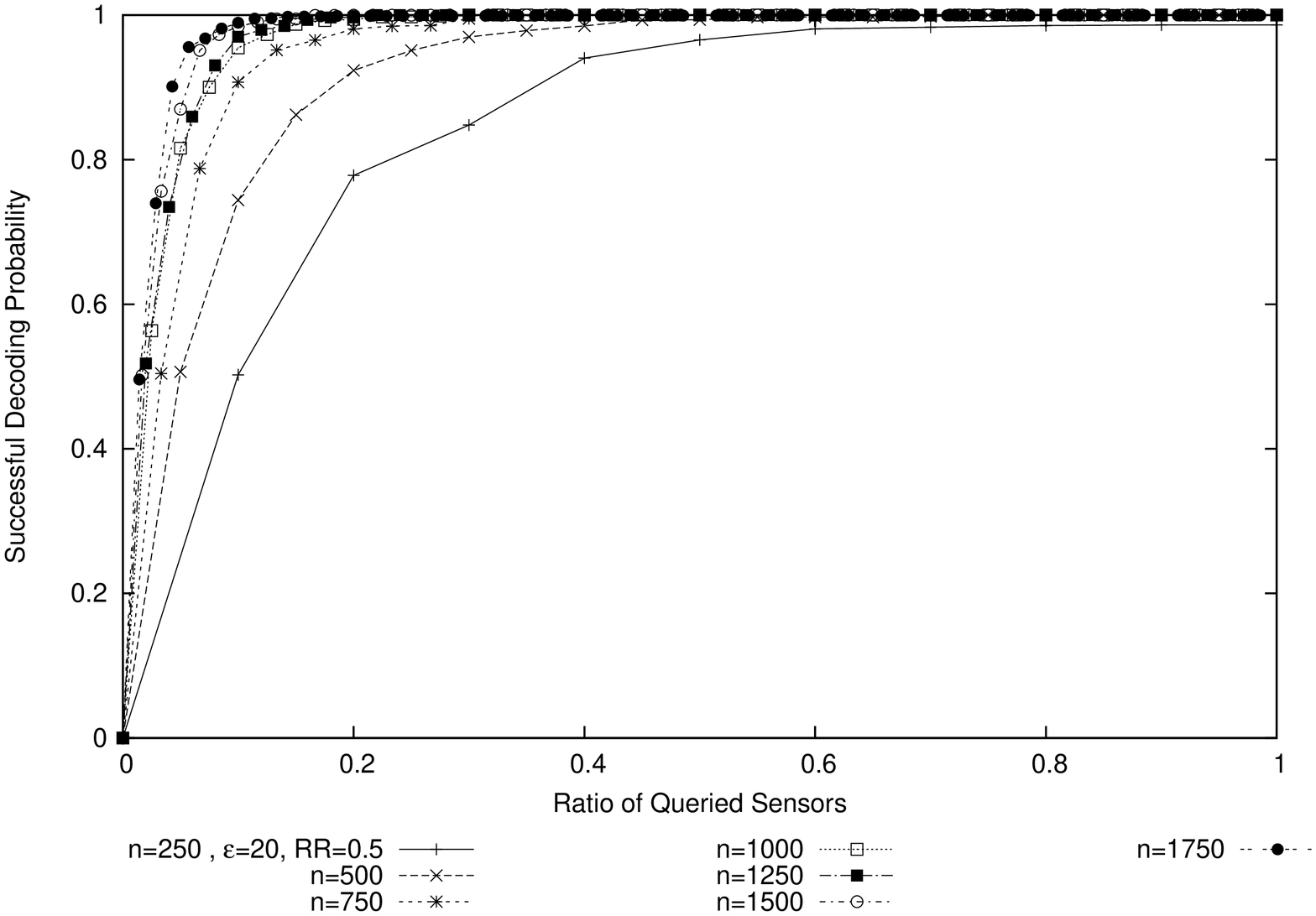}
  \caption{Network model representing  a wireless sensor network with sensing  and storage nodes. The successful decoding probability increases with increasing the total number of network nodes.}\label{fig:QueryRatioBuf160RR10}
  \end{center}
\end{figure}

Fig.~\ref{fig:QueryRatioBuf160RR10} shows that increasing the number of network nodes and fixing the covering radius of each node will result in an improvement in the successful decoding probability as well. Particularly, for $n>500$ and $n'>100$, we see that querying up to $20\%\sim 30\%$ will reveal the sensed data about all $k$ sensor nodes.
 Fig.~\ref{fig:QueryRatioBuf160RR10} shows that the revealed sensors ratio will be the same and approximately equals one when $\eta$ is greater than $0.17$ and also shows that for large number of sensors we will have larger $\rho$.

\begin{figure}[t]%
\includegraphics[scale=0.33,angle=270]{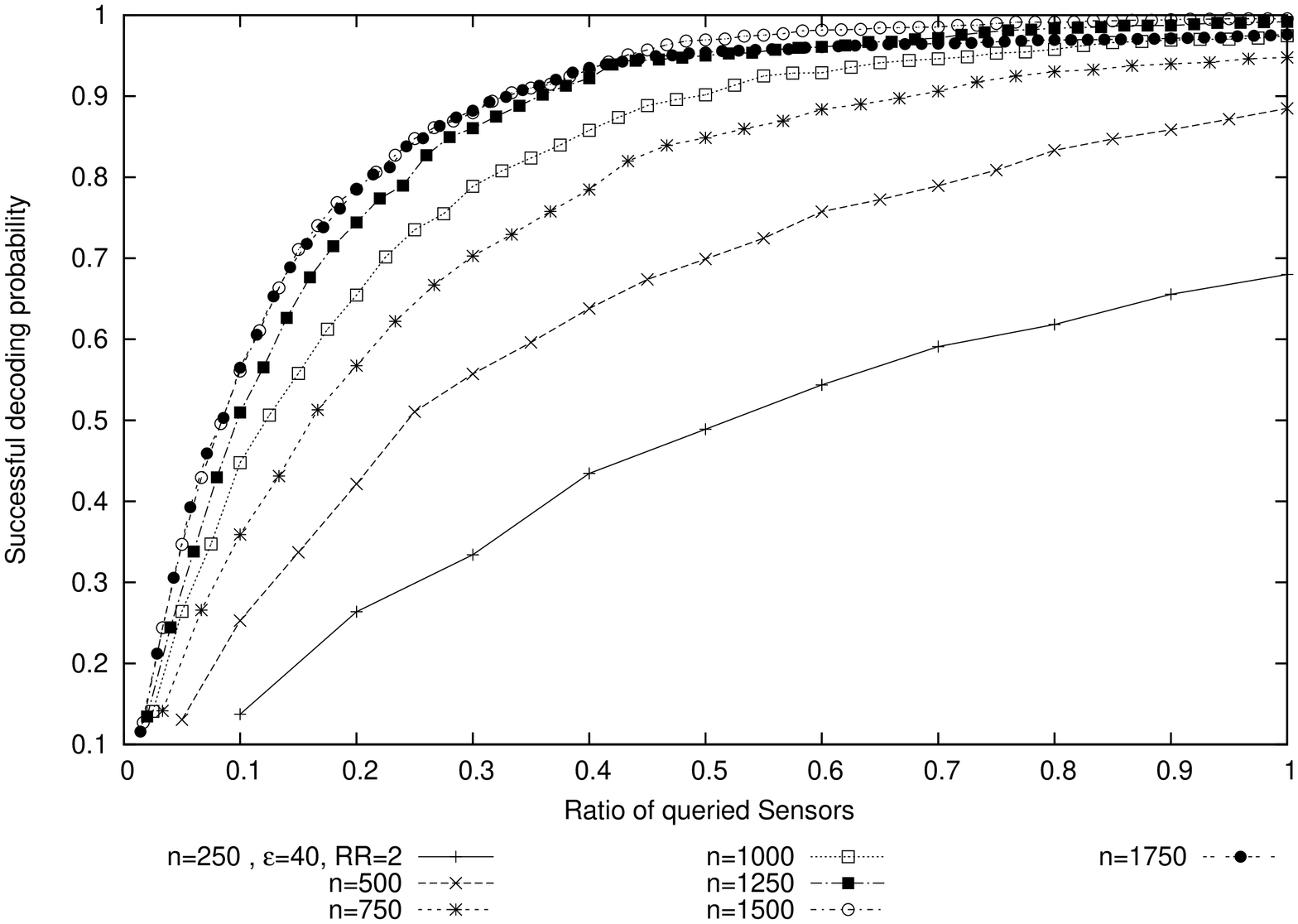}%
\caption{Effect of changing the percentage of queried nodes when the number of buffers and the radio range of all the nodes are changed.  Clearly increasing the number of nodes decreases the decoding performance due to lack of resources.}%
\label{fig:QuerR2B40}%
\end{figure}

In Fig.~\ref{fig:QuerR2B40}, we show the effect of increasing  the percentage of queried nodes on the successful decoding probability when each storage node has $40$ buffers and a radio range of $2$ distance units in this case of  a  square terrain of side length $L=100$ distance units. The percentage of storage nodes is always $20\%$ of the total number of nodes. Increasing the number of nodes has a positive effect on the successful decoding probability.  When there are $250$ total nodes, the nodes are more dispersed and with this small radio range, the storage nodes cannot reach all the sensor nodes and thus we are not able to decode more than $60\%$ of the sensors' data. This can be improved if the radio range is increased, thereby allowing the storage nodes to contact more sensors.

\begin{figure}[t]%
\includegraphics[scale=0.32,angle=270]{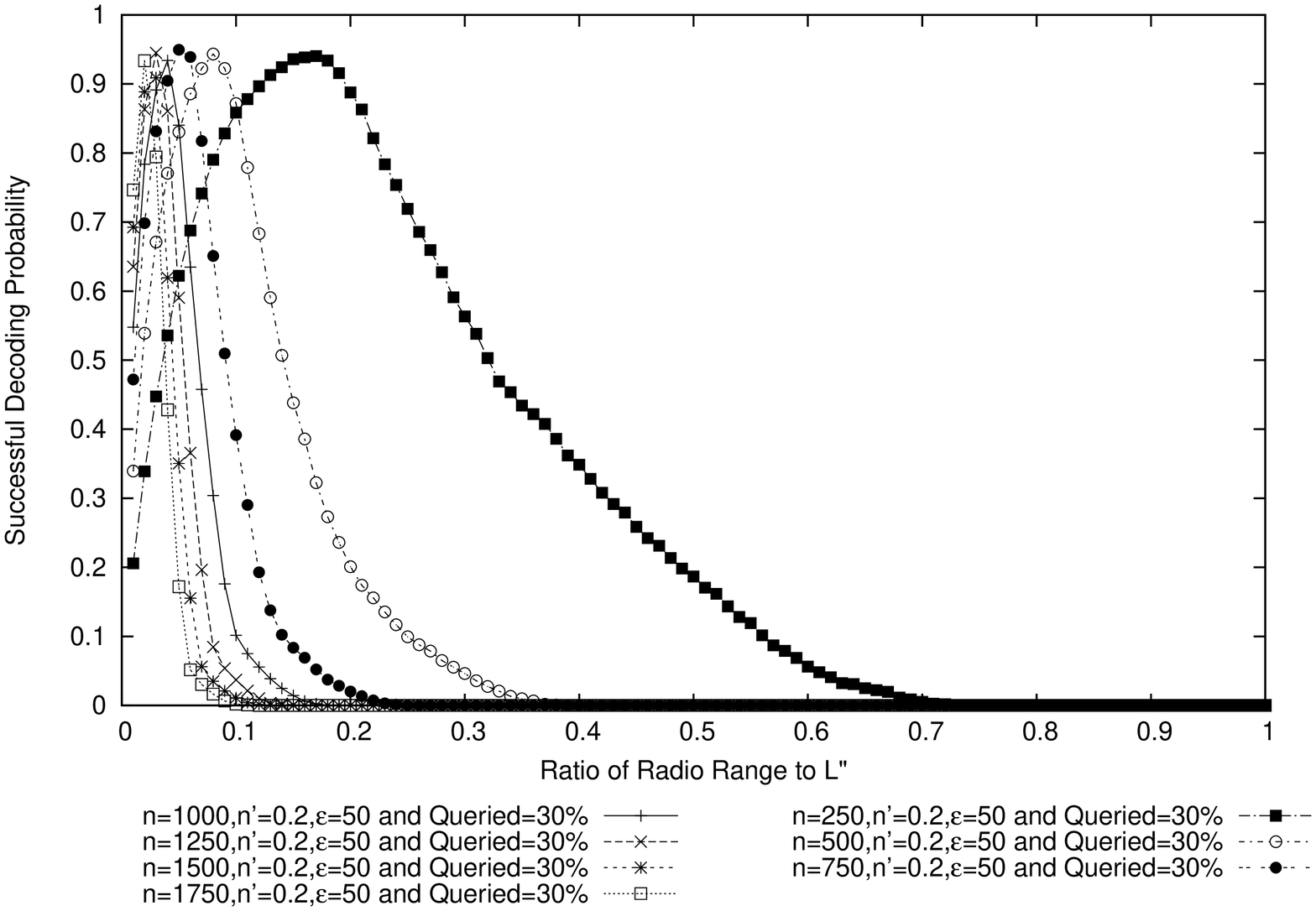}%
\caption{The effect of increasing the radio range with respect to $L$. The maximum radio range for better decoding performance depends on the number of nodes in the system.}%
\label{fig:RRBuf50}%
\end{figure}

Fig.~\ref{fig:RRBuf50} shows the effect of increasing the radio range with respect to the terrain side length $L$ when the buffer size can hold $50$ sensor messages per storage node and $30\%$ of the storage nodes are queried. As we increase the radio range, the number of encoded messages is increased. This makes decoding a much harder task until, at some point, no messages are decoded.  When the number of nodes in the terrain is limited, $250$ for example, the increase in the radio range results in an increase in the contacted nodes. This goes on until the radio range covers an area with a radius of almost $20\%$ of the side length of the terrain area. Increasing the radio range further results in encoding more nodes that we are not able to decode and thus in a gradual decrease of the successful decoding probability.  We can deduce from the curve that there is an optimal radio range for a network with a constant buffer size and node distribution beyond which the successful decoding probability decreases.

The simulation results demonstrate that the proposed model  is suitable for large-scale wireless sensor networks. Finding practical applications and network topologies in which this data collection algorithm can be deployed are directions for our future work.

%%%%%%%%%%%%%%%%%%%%%%%%%%%%%%%%%%%%%%%%%%%%%%%%%%%%%%%%%%%%%%%%%%%%%%%%
\section{Related Work} \label{sec:relatedwork}
In this section, we review some previous work in distributed data collection which is
relevant to our work.

%\smallskip

\begin{itemize}
\item

Dimakis~\emph{et al.} in~\cite{dimakis06a} and~\cite{dimakis06b} used a decentralized
implementation of fountain codes that uses geographic routing and every node
has to know its location. The motivation for using fountain codes instead of
using random linear codes is that the former requires $O(k \log k)$ decoding
complexity but the later such as  RS codes requires $O(k^3)$ decoding
complexity in which $k$ is the number of data blocks to be encoded.
%\smallskip

\item  Lin \emph{et al.} in~\cite{lin07b} and~\cite{lin07a} studied the question "how can we retrieve historical data that the sensors have gathered even
if some sensors are destroyed or disappeared from the network?"  They analyzed
techniques to increase "persistence" of sensed data in a random wireless sensor
network. They proposed two decentralized algorithms using fountain codes to
guarantee the persistence and reliability of cached data on unreliable sensors.
They used random walks to disseminate data from a sensor (source) node to a set
of other storage nodes. The first algorithm introduces lower overhead than
naive random-walk, while the second algorithm has lower level of fault
tolerance than the original centralized fountain code, but consumes much lower
dissemination cost.

\item
Kamara \emph{et al.} in~\cite{kamra06} proposed a novel technique called
\emph{growth codes} to increase data persistence in wireless sensor networks,
i.e. increasing the amount of information that can be recover at the sink.
\emph{Growth codes} is a linear technique that information is encoded in an
online distributed way with increasing degree. They defined persistence of  a
sensor network as \emph{"the fraction of data generated within the network that
eventually reaches the sink"}~\cite{kamra06}. They showed that \emph{growth
codes} can increase the amount of information that can be recovered at any
storage node at any time period.
%\medskip

\item
Aly~\emph{et al.} in~\cite{aly09f,aly08e} and \cite{aly10c} studied a model for distributed network storage algorithms for wireless sensor networks where $k$ sensor nodes (sources) want to disseminate their data to $n$ storage nodes with less computational complexity. The authors used fountain codes and random walks in graphs to solve this problem. They also assumed that the total numbers of sources and storage nodes are not known.

\end{itemize}

\section{Conclusion}\label{sec:conclusion}
In this paper, we have studied the distributed storage problem  in
large-scale random wireless sensor networks, in which there are sensing and storing nodes uniformly  distributed in a region. We have proposed a data collection algorithm to precisely collect sensed data and successfully store it at  storage nodes. The simulation results show that, with high probability, querying only $30\%$ of the storage nodes with limited or unlimited buffers will retrieve all sensed data gathered by the sensing nodes.

\bibliographystyle{plain}

\end{document}